\documentclass[sigconf, 10pt]{acmart}

\usepackage{multirow}
\usepackage{tcolorbox}
\usepackage[hypcap=false]{caption}
\usepackage{float}
\usepackage{hyperref}
\hypersetup{colorlinks,allcolors=black}
\usepackage{booktabs}
\usepackage{subcaption}
\usepackage{graphicx}
\usepackage{dblfloatfix} % Allows bottom placement of two-column floats

\renewcommand\footnotetextcopyrightpermission[1]{} % removes footnote with conference info
\setcopyright{none}

\acmDOI{}

\acmISBN{}

\acmConference[Submitted for Review]{}
\acmYear{2024}
\copyrightyear{}
\acmPrice{}
\begin{document}
\sloppy

\title{Towards Explainable Network Intrusion Detection using Large Language Models}
\settopmatter{authorsperrow=2}
\author{Paul R. B. Houssel} 
\email{p.houssel@uq.edu.au}
\affiliation{%
  \institution{The University of Queensland}
}

\author{Priyanka Singh} 
\email{priyanka.singh@uq.edu.au}
\affiliation{%
  \institution{The University of Queensland}
}

\author{Siamak Layeghy} 
\email{siamak.layeghy@uq.net.au}
\affiliation{%
  \institution{The University of Queensland}
}

\author{Marius Portmann} 
\email{marius@ieee.org}
\affiliation{%
  \institution{The University of Queensland}
}

% The default list of authors and the title is too long for headers}
\renewcommand{\shortauthors}{Houssel.et al.}
\renewcommand{\shorttitle}{Exploring LLMs for NIDS}

\begin{abstract}
Large Language Models (LLMs) have revolutionised natural language processing tasks, particularly as chat agents. However, their applicability to threat detection problems remains unclear. This paper examines the feasibility of employing LLMs as a Network Intrusion Detection System (NIDS), despite their high computational requirements, primarily for the sake of explainability. Furthermore, considerable resources have been invested in developing LLMs, and they may offer utility for NIDS. Current state-of-the-art NIDS rely on artificial benchmarking datasets, resulting in skewed performance when applied to real-world networking environments. Therefore, we compare the \textit{GPT-4} and \textit{LLama3} models against traditional architectures and transformer-based models to assess their ability to detect malicious NetFlows without depending on artificially skewed datasets, but solely on their vast pre-trained acquired knowledge. Our results reveal that, although LLMs struggle with precise attack detection, they hold significant potential for a path towards explainable NIDS. Our preliminary exploration shows that LLMs are unfit for the detection of Malicious NetFlows. Most promisingly, however, these exhibit significant potential as complementary agents in NIDS, particularly in providing explanations and aiding in threat response when integrated with Retrieval Augmented Generation (RAG) and function calling capabilities.
\end{abstract}

\maketitle

\section{Introduction}
Large Language Models (LLMs) have recently emerged as a transformative force in natural language processing, showcasing proficiency in handling unstructured data. With the ability to generate text and comprehend context, LLMs have found widespread application in various domains, ranging from conversational chat agents to code generation and translation tasks. Despite these advances, the application of LLMs to cybersecurity, particularly Network Intrusion Detection Systems (NIDS), remains largely unexplored.

NIDS are critical for monitoring and analysing network traffic to detect malicious activities and security breaches. Existing NIDS employ a combination of signature-based and anomaly-based detection techniques. Signature-based methods rely on predefined patterns of known attacks, while anomaly-based methods identify deviations from established norms in network behaviour. Although deep learning-based NIDS have demonstrated near-perfect performance on benchmark datasets, their efficacy in real-world environments is often questioned due to the prevalence of synthetic and misrepresentative datasets used in academic research~\cite{layeghy_benchmarking_2021}. Additionally, these models lack explainability~\cite{neupane_explainable_2022}, making it difficult for security professionals to understand and trust their predictions and respond to the detected threat. It thus remains important to investigate if LLMs are a potential solution to these.

Several studies have used transformers and LLMs to directly detect threats or as complementary solutions for NIDS. However, no previous work has thoroughly investigated LLMs' adaptivity to the domain of NetFlows. Previous studies replace LLM's sequence-to-sequence layers with a classification head, thus losing the potential for using the same model to provide explanations. Using LLMs and their natural language output, both detection by classification and explainability can be achieved. It remains to be seen if performance is degraded when using these text prediction models as threat classifiers, a use case very different from their original design purpose. LLMs have shown the ability to analyse long-term dependencies and be adaptive to the applied domains~\cite{xu_large_2024}. Furthermore, their pre-training on large amounts of natural language data might incorporate the knowledge required to detect malicious NetFlows without relying on skewed artificial datasets. This paper explores the capability of LLMs to detect network attacks based on network flows while contrasting their benefits and limitations with traditional Machine Learning (ML) solutions.

We empirically assess the pre-trained OpenAI's\textit{GPT-4} and Meta's\textit{LLama3} models, in the context of zero-shot learning and fine-tuning on standardised NetFlow datasets. Our initial experiments investigate the performance capabilities of LLMs to detect threats, a trait which traditional more lightweight ML solutions have already been proven to possess with near perfection~\cite{sarhan_netflow_2021}. We will evaluate our model variants on the \textit{NF-UNSW-NB15-v2} and \textit{NF-CSE-CIC-IDS2018-v2} datasets~\cite{sarhan_towards_2022}. Finally, we investigate the capacity to explain the detection of threats and their computational cost of inference to fuel a discussion on the advantages and drawbacks of using LLMs for NIDS versus more lightweight solutions incorporating LLMs as a complementary solution to existing NIDS.
Our findings aim to provide a first exploratory evaluation of LLMs in the context of NIDS, offering insights into potential future research directions and use cases. We do not aim for a comprehensive benchmark of models and parameters of LLMs for detecting malicious NetFlows.

\section{Related Work}
\subsection{Malicious NetFlow Detection}
In contrast to NIDS, Host-based Intrusion Detection Systems (HIDS) typically deal with textual data such as system calls, logs, or application and memory traces. Natural Language Processing (NLP) techniques are employed to examine the semantic interactions and sequences of system calls and contextual data, enabling the effective identification of suspicious activities. This facilitates precise, real-time intrusion detection by analysing text data streams. Many solutions relying on NLP have been developed even before the advent of LLMs~\cite{sworna_nlp_2022}. Natural language is less present in the NIDS domain, but surprisingly more literature exists.

For detection based on network traffic, the application of LLMs has started to gain traction, although the solutions often modify the standard LLM architecture. Specifically, these models typically strip the sequence-to-sequence head and replace it with a classification head to obtain a deterministic output to evaluate the model. The initial textual output head, which predicts the next textual token, is replaced by a classification head vector which has a dimension equivalent to the number of prediction classes, \textit{e.g} two for a binary classification. Lira et al.~\cite{g_lira_harnessing_2024} investigated the capabilities of LLMs such as \textit{GPT-3.5}, \textit{GPT-4}, and \textit{ADA} to detect Distributed Denial of Service (DDoS) attacks on the \textit{CICIDS2017}~\cite{sharafaldin_toward_2018} and \textit{Urban IoT}~\cite{hekmati_correlation-aware_2023} datasets. Their findings indicate that LLMs, when fine-tuned or employed in few-shot learning scenarios, could detect DDoS attacks with significant accuracy proving their adaptivity trait. Manocchio et al.~\cite{manocchio_flowtransformer_2024} explored various encoding and classification methods on different transformer architectures, including \textit{GPT}'s and \textit{BERT}'s to enhance network analysis. Ferrag et al.~\cite{ferrag_revolutionizing_2024} used \textit{BERT} for threat detection in Internet of Things (IoT) networks, outperforming existing ML solutions by distinguishing between 14 attack classes and benign traffic using \textit{PCAP} files. Alkhatib et al.~\cite{alkhatib_can-bert_2022} explored the usage of \textit{BERT} for detecting threats on Controller Area Networks (CAN), showcasing the model's versatility across different cybersecurity applications. Among these, Chen et al.~\cite{guastalla_application_2024} is the only study which employed LLMs, which are pre-trained on natural language and possess a sequence-to-sequence head producing sequences of natural text tokens, to detect DDoS attacks by fine-tuning \textit{BERT}.

\subsection{Explainable NIDS}
Studies aiming to render NIDS explainable typically employ existing explainable AI methods. These methods assign an importance factor to each input feature, indicating their influence on the model's output label~\cite{neupane_explainable_2022}. For instance, Mallampati et al. \cite{mallampati_enhancing_2024}, Senevirathna et al.~\cite{senevirathna_deceiving_2024}, and Zebin et al.~\cite{zebin_explainable_2022} utilise the SHapley Additive exPlanation (SHAP) method across various network domains. This approach expresses explainability as the weight of each feature and its impact on the output class, enabling the development of algorithms to respond to threats in real time. Wei et al.~\cite{wei_xnids_2023} propose a framework that leverages this feature's importance to design rules for active intrusion responses. Their XNIDS framework identifies features critical to the prediction class and generates actionable defence rules, which aid in understanding DL-NIDS behaviours and troubleshooting detection errors.

LLMs could potentially provide natural language explanations for human operators, not only limiting the explanations to features which statistically appear anomalous to the NIDS. This will allow to facilitate threat response and a better understanding of their NIDS functionality or potential malfunctioning. In this study, we further investigate the capability of LLMs to detect malicious NetFlows while explaining their reasoning. These could potentially be employed as a complementary solution, aiming to enhance the explainability of NIDS. 

\section{Methodology}
This section outlines the methodology employed to investigate the potential of LLMs for NIDS. Our approach is based on the open-source \textit{LLama3-8B-Instruct} and \textit{GPT-4} model, designed to better handle instruction-based tasks compared to the foundational model. In contrast with foundational model's difficulties in answering in the required format, as it was not trained to understand instructions but simply to predict sequences of words. We focus on the network traffic under the format of flows, which are currently the most prominent and efficient source of network telemetry for threat detection due to their scalability~\cite{layeghy_benchmarking_2021}. They summarise network traffic which shares destination and source port and IP addresses by aggregating packets into flows, making them ideal for detecting patterns indicative of malicious activities. Sarhan et al.~\cite{sarhan_towards_2022} proposes a set of standardised NetFlow datasets, with a common set of overlapping features, to address the limited ability to evaluate ML-based NIDS across different network environments and attack scenarios. Among these, we solely consider the datasets outside of the IoT domain: \textit{NF-UNSW-NB15-v2} and \textit{NF-CSE-CIC-IDS2018-v2}.

\subsection{Domain Specific Performance}
To measure the performance on specific network environments, we built two baseline models using the \textit{LLama3-8B-Instruct} and \textit{GPT-4-0613} model by applying the zero-shot learning paradigm. Additionally, we will investigate two fine-tuning methods only on the \textit{LLama3} model due to the inherited cost associated with fine-tuning using the \textit{OpenAI} API. For this series of experiments, all datasets are split into training and testing sets with a 95\% to 5\% ratio while stratifying the attack type. All evaluations are performed following a 10-fold cross-validation method. We only consider a small testing set due to the sheer computational cost of LLM inference. All prediction results from our experiments are evaluated using the macro average of Precision and Recall for both the benign and malicious classes. The macro average is the unweighted sum of the individual performance scores for each class, representing each class equally. We provide the pre-trained LLM with an instruction to detect network attacks based on NetFlows, requesting it to output either \textit{"1"} or \textit{"0"} to indicate whether the provided flow is malicious or benign. After ingesting the instruction prompt, the model will take as an input a tabular NetFlow entry encoded as text using key-value pairs separated by commas to represent the feature name and value pairs. For instance, a network flow originally represented as a row within the CSV table of the dataset is transformed into text as follows: "\textit{L4\_DST\_PORT: 80}". To ensure consistency and reduce variability in outputs, we set the model's temperature parameter to $10^{-1}$. The temperature parameter, ranging from 0 to 1, influences the probabilities generated by the softmax output function. A low-temperature setting makes the model's outputs more deterministic by prioritising the most likely tokens, important for binary classification tasks~\cite{minaee_large_2024}. Together with prompt engineering, it allows us to constrain the response of the model to a binary output without any exceptions.

\paragraph{\textbf{Zero-shot Learning}}

Zero-shot learning consists of a pre-trained model tasked with making predictions on yet explicitly unseen data. In the context of a pre-trained LLM, zero-shot learning leverages the knowledge and understanding embedded in the model from its training on a vast amount of natural language. When applied to zero-shot learning, the LLM can make informed predictions to identify malicious network traffic by understanding the provided instructions and applying its broad contextual knowledge to classify the traffic as either malicious or benign, even though it hasn't been specifically trained on labelled examples of network attacks.

\paragraph{\textbf{Fine-tuning}}
Fine-tuning LLMs for specific tasks is a crucial step to enhance their performance and adapt them to domain-specific requirements. In our investigation, we focus on two state-of-the-art fine-tuning methods: Odds Ratio Preference Optimisation (ORPO)~\cite{hong_orpo_2024} and Kahneman-Tversky Optimisation (KTO)~\cite{ethayarajh_kto_2024}. These methods have shown significant promise and represent the current state of the art~\cite{minaee_large_2024}. ORPO~\cite{hong_orpo_2024}, requires preference data in the format of prompts paired with rejected and accepted answers. For the task of NIDS binary classification, we can adapt these methods by treating the true label as the accepted prompt and the inverse of the label as the rejected prompt. In contrast, KTO~\cite{ethayarajh_kto_2024} does not require paired preference data and instead relies on a series of prompts and corresponding responses along with a series of binary labels indicating wherever the provided response is relevant. In our case, we use the true label as the accepted response. Half of the prompts are randomly sampled with relevant responses using the label and the other half are samples as irrelevant by using the inverse of the label.

\subsection{Explainability}
To investigate and grasp the LLM's ability to explain the reasons for predicting a network flow as benign or malicious, we manually analyse the models' responses to find patterns and inconsistencies. More specifically, we analysed the explanations provided by our baseline model for the cases of True Positives (TP), False Positives (FP), True Negatives (TN), and False Negatives (FN) samples, as these will offer variability in answering. This analysis is not a quantitative evaluation of the quality of explanations but serves as an initial exploration to assess the importance of future research into making LLMs explainable for NIDS.

\subsection{Computational Complexity}
To assess the computational cost, we compared our baseline with the most common architectures with the default parameters, used in related work and industrial application of NIDS. All experiments were conducted on a machine equipped with an NVIDIA RTX3090 GPU and a 3.50 GHz CPU. The measured time for sequential inference corresponds to the average time taken to predict a single NetFlow sample. This time was measured by calculating the average time to predict a single sample when processing the entire testing set as a batch, computing the average over 10 runs.

\section{Empirical Results}

\subsection{Domain Specific Performance}

\paragraph{\textbf{Zero-shot Learning}}
The results of these empirical experiments under the zero-shot evaluation (Table \ref{tab:domain-specific-performance-1} and \ref{tab:domain-specific-performance-2}) reveal several important insights. Unsurprisingly, the performance of the pre-trained \textit{LLama3} model is worse than random selection, which would have been, considering the macro average, 50\% for all metrics, while \textit{GPT-4} is slightly better. This outcome is expected because pre-trained LLMs are primarily designed for natural language tasks and not for the specific task of network classification. In contrast, classical ML models demonstrate near-perfect performance. For instance, on the \textit{NF-UNSW-NB15-v2} dataset, a Random Forest and an LSTM neural network respectively achieve a weighted F1-score of 92.17\% and 92.82\%~\cite{layeghy_explainable_2023} while a \textit{DANN} model achieves 97.81\%~\cite{layeghy_di-nids_2023}.

\begin{table}[h!]
    \centering
    \caption{Domain-specific zero-shot performance of the model on the \textit{NF-UNSW-NB15-v2} dataset.}
    \begin{tabular}{cccc}
    \hline
    \toprule
    \textbf{Model} &\multicolumn{2}{c}{\textbf{Macro Avg. Perf. Metrics$\uparrow$}}\\
    &Precision & Recall\\
    \midrule
    \textit{LLama3-8b-Instruct}&48.56\%&40.56\%\\
    \hline
    \textit{GPT-4}&50.85\%&53.47\%\\
    % \multirow{3}{*}{\textit{KTO} Finetuned}&1,000&52.69\%&50.22\%&49.60\%\\
    % &10,000&53.98\%&50.36\%&49.88\%\\
    % &50,000&55.18\%&50.67\%&50.48\%\\
    % \hline
    % \multirow{3}{*}{\textit{ORPO} Finetuned}&1,000&48.88\%&42.76\%&36.96\%\\
    % &10,000 &49.25\% &40.40\%&44.35\%\\
    % &50,000 &54.00\% &43.50\%&48.17\%\\
    \bottomrule
    \end{tabular}
    \label{tab:domain-specific-performance-1}
\end{table}
\begin{table}[h!]
    \centering
    \caption{Domain specific zero-shot performance on the \textit{NF-CSE-CIC-IDS2018-v2} dataset.}
    \begin{tabular}{cccc}
    \hline
    \toprule
    \textbf{Model}& \multicolumn{2}{c}{\textbf{Macro Avg. Perf. Metrics$\uparrow$}}\\
    &Precision & Recall\\
    \midrule
    \textit{LLama3-8b-Instruct}&49.66\%&49.73\%\\
    \midrule
    \textit{GPT-4}&50.25\%&51.02\%\\
    % \multirow{3}{*}{\textit{KTO} Finetuned}&1,000&44.02\%&49.99\%&46.82\%\\
    % &10,000&44.02\%&45.00\%&46.83\%\\
    % &50,000&44.02\%&50.00\%&46.82\%\\
    % \hline
    % \multirow{3}{*}{\textit{ORPO} Finetuned}&1,000&39.43\%&24.95\%&28.48\%\\
    % &10,000&40.00\%&29.01\%&33.10\%\\
    % &50,000&42.50\%&35.00\%&38.03\%\\
    \bottomrule
    \end{tabular}
    \label{tab:domain-specific-performance-2}
\end{table}

\paragraph{\textbf{Fine-tuning}}
In comparison to the zero-shot learning results, both fine-tuning methods showed slight improvements but still failed to significantly outperform random selection or approach the performance of related works (Table.~\ref{tab:domain-specific-performance-fine-tuning-1} and \ref{tab:domain-specific-performance-fine-tuning-2}). Overall, \textit{KTO} performed slightly better, indicating that the choice of fine-tuning method is an important parameter for future research. We see that fine-tuning can enhance performance with increased training, but the improvements observed were minimal. Fine-tuning tends to adjust the model's responses to fit a desired style but is insufficient to significantly alter the weights for the binary classification of network attacks. Pre-trained natural language models, which are trained on extensive natural language corpora, may not be capable of recognising the subtle patterns that indicate malicious NetFlows. They are more adept at generating appropriate responses to natural language queries without any further deep understanding.

Fine-tuning LLMs for detecting malicious NetFlows is unnecessary; instead, transformer-based models can be trained directly on entire datasets. Manocchio et al., through their proposed \textit{FlowTransformer} framework~\cite{manocchio_flowtransformer_2024}, have shown that transformer architectures perform well on binary NetFlow classification tasks when trained on complete datasets. This performance is achieved without the need pre-training on natural language datasets. For instance, their model, using GPT's architecture achieves an 89.98\% F1-score.
\begin{table}[h!]
    \centering
    \caption{Domain specific performance of the \textit{LLama3-8b-Instruct} model on the \textit{NF-UNSW-NB15-v2} dataset.}
    \resizebox{0.5\textwidth}{!}{
    \begin{tabular}{ccccc}
    \hline
    \toprule
    \textbf{Learning} & \textbf{\# Training}& \multicolumn{2}{c}{\textbf{Macro Avg. Perf. Metrics$\uparrow$}}\\
    \textbf{Paradigm}&\textbf{Samples}& Precision & Recall\\
    \midrule
    zero-shot&0&48.56\%&40.56\%\\
    \hline
    \multirow{3}{*}{\textit{KTO} Finetuned}&1,000&52.69\%&50.22\%\\
    &10,000&53.98\%&50.36\%\\
    &50,000&55.18\%&51.67\%\\
    \hline
    \multirow{3}{*}{\textit{ORPO} Finetuned}&1,000&48.88\%&40.76\%\\
    &10,000 &49.25\% &42.40\%\\
    &50,000 &52.00\% &43.50\%\\
    \bottomrule
    \end{tabular}}
    \label{tab:domain-specific-performance-fine-tuning-1}
\end{table}

\begin{table}[h!]
    \centering
    \caption{Domain specific performance of the \textit{LLama3-8b-Instruct} model on the \textit{NF-CSE-CIC-IDS2018-v2} dataset.}
    \resizebox{0.5\textwidth}{!}{
    \begin{tabular}{cccccc}
    \hline
    \toprule
    \textbf{Learning} & \textbf{\# Training}& \multicolumn{2}{c}{\textbf{Macro Avg. Perf. Metrics$\uparrow$}}\\
    \textbf{Paradigm}&\textbf{Samples}& Precision & Recall\\
    \midrule
    zero-shot& 0&49.66\%&49.73\%\\
    \hline
    \multirow{3}{*}{\textit{KTO} Finetuned}&1,000&50.02\%&50.09\%\\
    &10,000&52.22\%&55.13\%\\
    &50,000&55.02\%&58.79\%\\
    \hline
    \multirow{3}{*}{\textit{ORPO} Finetuned}&1,000&50.43\%&50.95\%\\
    &10,000&50.89\%&51.01\%\\
    &50,000&53.60\%&51.58\%\\
    \bottomrule
    \end{tabular}}
    \label{tab:domain-specific-performance-fine-tuning-2}
\end{table}

\newpage

\subsection{Explainability}
The analysis of LLama3 generated explanations for various
prediction cases reveal unique benefits and notable limitations
to enrich future NIDS applications with explainability,
a unique feature proper to LLMs. These sample prompts
demonstrate that LLMs can provide detailed and contextually
relevant explanations for their classifications of network flows
but with notable limitations.

When asked about a TN sample (Fig.~\ref{explaination-tn}), the model correctly identifies the network flow as a DNS query. However, it inaccurately claims the source IP address is from China, whereas it is actually from Japan, and misidentifies the destination address as being in the USA, although it belongs to the University of New South Wales (UNSW), the dataset's creator. Despite these errors, the LLM correctly identifies the byte size range of all packets in the flow as typical for a DNS query. These inaccuracies highlight the LLM's tendency to hallucinate or generate incorrect information, particularly regarding geographical locations.

On a TP case, the LLM mistakenly states that protocol 139 corresponds to the \textit{NetBIOS} protocol, whereas it corresponds to the \textit{Host Identity Protocol}. Its confusion is evident as NetBIOS uses port 139. Most importantly, it operates at Layer 5, while the \textit{protocol} feature identifies those of Layer 4. Additionally, the LLM incorrectly assumes that the use of ports with the value 0 indicates malicious activity. While unusual, ports set to 0 can occur in protocols not using port numbers, like \textit{ICMP}. These errors show that the LLM has gaps in its understanding of protocol specifics and port usage, leading to incorrect conclusions about malicious behaviour.

In the FP scenario, unusual patterns, such as atypical ports and large file transfers are identified leading to the conclusion that these imply maliciousness. While the reasoning is plausible, it cannot correlate these features to specific attack signatures accurately.

For the FN sample, the model's explanation is mostly consistent. However, it incorrectly identifies the main Layer 7 protocol user and mistakenly states that Layer 7 protocol number 7 corresponds to HTTP when it indicates the Internet Printing Protocol (IPP). The model classifies the network flow as benign due to small packet lengths and a low number of retransmitted packets, correctly identifying the traffic as typical HTTP traffic. While the model can accurately explain benign traffic by recognising common patterns, it fails to identify the malicious aspects of the sample.

% \begin{tcolorbox}[colback=blue!5!white,colframe=blue!75!black]
%      \small It appears to be a typical HTTP request from a client (source IP address) to a web server (destination IP address). Here are some reasons that led me to make this classification:
%      \begin{itemize}
%          \item Destination port: The destination port is 80, which is the standard port for HTTP.
%         \item L7 protocol: The L7 protocol is set to 7.0, which corresponds to HTTP.
%         \item TCP flags: The TCP flags are 19, which indicates that the connection was established and data was exchanged in both directions.
%         \item Packet lengths: The packet lengths are relatively small, ranging from 40 to 269 bytes, which is consistent with typical HTTP requests and responses.
%         \item Retransmitted packets: There are a few retransmitted packets, but this is not unusual for TCP connections.
%      \end{itemize}
% Overall, the flow exhibits characteristics that are commonly seen in legitimate web traffic, such as HTTP requests and responses.
% \end{tcolorbox}
% \noindent\begin{minipage}{.45\textwidth}
% \captionof{figure}{Explanation Prompt for a False Negative Netflow.}\label{explaination-fn}
% \end{minipage}
\begin{tcolorbox}[colback=blue!5!white,colframe=blue!75!black]
 \small
    \begin{itemize}
        \item  Protocol: The protocol used is UDP (17), which is commonly used for DNS queries. This suggests that the flow may be related to a legitimate DNS request.
        \item Destination port: The destination port is 53, which is the standard port for DNS traffic.
        \item Source and destination addresses: The source address is an IP address in China, while the destination address is an IP address in the United States. This could indicate that the flow is related to a legitimate DNS query from a Chinese user to a US-based DNS server.
        \item Packet sizes and lengths: The packet sizes are relatively small (73-89 bytes), which is typical for DNS.
        \item Flow duration: The flow duration is 0 milliseconds, indicating that this was likely a single, brief request-response exchange.
    \end{itemize}
\end{tcolorbox}
\noindent
\begin{minipage}{0.47\textwidth}
\captionof{figure}{Explanation provided for a True Negative NetFlow.}
\label{explaination-tn}
\end{minipage}

Overall, from these manual experiments, we learned that LLM's explanations are convincing and correctly retrieve information from the NetFlow data, being consistent with the presented features. However, they often fail to reason logically and augment the information with factual accuracy. The LLM tends to puzzle feature values together to identify traffic types (e.g., \textit{HTTP} traffic, \textit{DNS} queries) effectively, which can be useful for network operators. However, it struggles to correlate multiple features to identify attack signatures accurately, instead treating individual suspicious features as independent clues of malicious activity. One significant problem is their tendency to generate nonsensical or unfaithful content, known as ``hallucination``~\cite{minaee_large_2024}. These models lack a true comprehension of facts or logical reasoning, relying on learned patterns of word succession within context. This is even more evident in further prompts. When questioned about the units of measurement used in network flows, the LLM failed to provide a coherent answer. When provided with a malformed NetFlow, the LLM was unable to determine its soundness. Additionally, the LLM failed to explain a SYN flood attack, a straightforward task for signature-based detectors like \textit{Cisco Snort}.

\subsection{Computational Complexity}

The feasibility of using LLMs for NIDS not only relies on detection performance but as importantly on inference time. Given that NIDS must handle hundreds, if not thousands, of NetFlows per second, these models must keep up with the network traffic and be able to report incidents promptly. Delays in inference can result in missed or late alerts, compromising network security.

\begin{table}
      \centering
      \caption{Complexity of benchmarked model compared to traditional NIDS architectures.}
      \resizebox{0.48\textwidth}{!}{
      \begin{tabular}{ccc}
            \hline
            \toprule
            \textbf{Method} &\textbf{Inference Time ($\mu s$)}& \textbf{\#Parameters}\\
            \midrule
            \textit{LLama3-8B-Instruct}&14,000.01&8,030,000,000\\
            \hline
            Random Forest&2.03&867,614 (nodes)\\
            % Feed Forward NN& &&\\
            Decision Tree~\cite{manocchio_network_2022}&1.53&--\\
            LSTM NN&25.02&56,555\\
            DANN &28.40&67,208\\
            % DI-NIDS &  &\\
            \bottomrule
          \end{tabular}}
      \label{tab:performance}
\end{table}

Our comparison in Table.~\ref{tab:performance}, highlights the significant difference in computational complexity among the models. The \textit{LLama3} model exhibited significantly higher inference times due to its complex architecture and a large number of parameters. As \textit{GPT-4} is closed-source and runs on the server side, we were unable to obtain statistics about its inference times and parameter count, which are in a similar range to those of \textit{LLama3}.
In contrast, traditional ML models like the Random Forest and the Decision Tree demonstrated swift inference times, making them more suitable for real-time NIDS applications. For instance, DL models can still offer a reasonable trade-off compromise, with costs potentially being manageable. However, considering the near 7000-fold slower inference time of LLMs compared to lightweight models, it is not feasible to implement LLMs in a real-time NIDS.

\section{Discussion}
Current fine-tuning approaches have not significantly improved performance, highlighting the need for more effective methods tailored to classification tasks.
Parameters such as temperature, data encoding and prompt formulation can also affect performance and need to be studied comprehensively for their impact on classification.
Furthermore, as demonstrated by the cross-domain performance limitations of existing NIDS models~\cite{layeghy_explainable_2023}, there is a need for more robust and adaptable architectures. Comparing LLMs, known for their adaptability, with other ML architectures in terms of their performance on unseen network behaviours could offer insights for future development in this field.

To improve explainability, the integration of a Retrieval-Augmented Generation (RAG) system~\cite{lewis_retrieval-augmented_2020}, linked with a Cyber Threat Intelligence (CTI) source, could significantly enhance the reliability of LLMs. RAG can ground LLMs in accurate, up-to-date information, potentially reducing or eliminating hallucinations. Expanding RAG beyond CTI to include specifications of observed systems, such as endpoints, operating systems, SBOM, and firewall rules, could enhance both detection rates and response effectiveness. On the other hand, to provide explainable threat response, deployed LLMs should be equipped with access to function calls. Function calling, also known as tool use or API calling, enables LLMs to interact with external systems~\cite{liang_taskmatrixai_2023}. By providing the LLM with a predefined set of functions or tools along with usage instructions, it can intelligently select and invoke appropriate functions to address a given threat scenario. For instance, it could propose specific functions with parameters to temporarily modify firewall rules. To evaluate the progress of future research in this area, it is essential to develop a methodology for quantitatively assessing the explainability of LLMs to reason on NIDS alerts. Currently, datasets necessary to fuel this methodology do not exist, highlighting future research.

While LLMs show promise as assistants for providing explanations, they require further research and development before they can be considered reliable in this capacity. Future work shall also focus on LLMs' understanding of network traffic, as it is crucial for their potential role as a virtual system administrator, complementing NIDS. This research would build upon the work of Donadel et al.~\cite{donadel_can_2024}, evaluating their understanding of network topology.
\section{Conclusion}
Our investigation into the application of LLMs for NIDS has revealed notable limitations in their performance as primary solutions. If used as a solution to detect threats based on NetFlows, future research shall rather focus on transformer-based architectures without relying on pre-trained LLMs. LLMs currently struggle to effectively detect domain-specific malicious NetFlows and have high inference time complexity, making them impractical as standalone NIDS solutions. Given these findings, we propose that LLMs should be harnessed as complementary solutions to existing state-of-the-art NIDS. Specifically, their strengths in providing explainability make them valuable when integrated with traditional ML-based systems, particularly for generating detailed and contextually relevant explanations when alerts are raised. This approach leverages the potential of LLMs to enhance the interpretability and transparency of NIDS, thus aiding in threat response and decision-making processes. While challenges such as hallucination and logical reasoning limitations persist, the promising results in explainability underscore the importance of further exploring LLMs to finally render NIDS explainable.
\newpage
\bibliographystyle{ieeetr}
\bibliography{references}

% \appendix
% \begin{tcolorbox}
% "As a cybersecurity expert, your task is to classify network flows as either malicious or benign. If you determine the network flow is benign, respond with '0'. If you determine the network flow is malicious, respond with '1'.

% Each NetFlow entry will be provided as key-value pairs representing the following features:

% \begin{itemize}
%     \itemsep 0em
%     \item IPV4 SRC ADDR: IPv4 source address
%     \item \textbf{[...]}
%     \item FTP COMMAND RET CODE: FTP client command return code
% \end{itemize}

% Respond only with '0' or '1'. If requested, explain your classification, detailing the reasoning and which feature values influenced your decision."
% \end{tcolorbox}
% \noindent\begin{minipage}{0.45\textwidth}
% \captionof{figure}{Instruction Prompt for the evaluation of \textit{NF-v2} datasets.}\label{prompt-1}
% \end{minipage}
\end{document}